\documentclass{article}
\usepackage{authblk}
\usepackage{gensymb}
\usepackage{graphicx}
\usepackage{amsmath}
\usepackage{amssymb}
\usepackage[numbers,square]{natbib}

\title{Morphology dependent surface properties of nanostructured GaN films grown by molecular beam epitaxy}

\author[1]{Abhijit Chatterjee}
\author[2]{Swathi S.P.}
\author[3*]{S.M. Shivaprasad}
\affil[1,2,3*]{Chemistry and Physics of Materials Unit, Jawaharlal Nehru Centre for Advanced Scientific Research, Jakkur, Bangalore-560064, India}

\begin{document}
\maketitle
\section{Introduction}
The group III-nitrides are excellent materials for fabrication of different types of electronic and optoelectronic devices \citep{ruterana_nitride_2006} owing to their superior electronic and structural properties such as wide, direct and tunable bandgap, high electron mobility, high electron peak velocity, stability in harsh chemical environments, high thermal conductivity, etc. Examples of applications of III-nitrides  include LEDs, lasers, high power and high frequency electronics, sensors, photovoltaics, etc.  However, to attain optimal performance of these devices several issues need to be resolved. Among the various limitations, forming dislocation and defect free III-nitride thin films, p-doping the material and making ohmic contacts to the device structures remain predominant, and are constantly addressed by researchers. Though several approaches have been adopted, of late, there is great interest in nanostructuring the material to exploit their novel properties in low dimensions, due to electron and photon confinement, strain relaxed growth and large surface to volume ratio.\citep{pearton_gan_2012}     

In the recent past,  we have reported several interesting results on the formation and properties of the self-assembled GaN nanowall network.\citep{kesaria_evidence_2011},\citep{bhasker_high_2012} This configuration displays  unprecedented properties such as low defect, high conductivity, high mobility, etc. Fabrication of  devices exploiting these nanostructures necessitates a thorough understanding of the surface and interface properties of different morphological manifestations of materials which display different surface chemistry and electronic states. This difference, in turn, affects formation of metal contacts, \citep{rickert_n-gan_2002} Fermi level pinning, \citep{schmitz_metal_1998} band bending \citep{tracy_preparation_2003} etc, all of which are crucial in device fabrication. By kinetically varying growth conditions, nanostructured films of various morphologies can be deposited. Density and type of point defects and dislocations, strain, etc, all depend on the structure and  morphology of the material and therefore electrical and luminescence properties can be tuned by changing the kinetics of growth. In this work we have grown GaN samples of different morphologies using PAMBE by controlling the growth parameters and probed their surface chemistry and electronic structure, and related them to the corresponding electrical and luminescence properties.

\section{Experimental}
The GaN thin films S1 to S3 were grown using a radio frequency plasma assisted molecular beam epitaxy system (SVTA). High purity Ga metal (99.99999 \%) and nitrogen gas (99.9995\%) were used as sources. C-plane sapphire ($\alpha$-$Al_{2}O_{3}$) was used as the substrate, which was degreased by ultrasonicating in organic solvents and rinsed in deionized water, before being inserted in the preparation chamber of MBE. Here, it was subjected to thermal cleaning by degassing it at 600 \degree C for 1 hour. The substrate is then transferred to the main growth chamber where it is  again degassed at 850 \degree C for 30 minutes. The samples S1 to S3 were grown in nitrogen rich conditions, while the substrate was held at   630 \degree C. Ga K cell temperatures of 1050 \degree C, 1020 \degree C, 1100 \degree C and nitrogen flow rates of 4.5 sccm, 6 sccm and 4.5 sccm were used for forming samples S1, S2 and S3, respectively. All samples were grown for a duration of 4 hours with the RF plasma forward power of 375 W. The sample S4 is a commercially obtained 2 $\mu$m thick MOCVD grown GaN flat epilayer which is used for comparative studies of morphology dependence of properties.  The morphology of the samples were observed by a Field Emission Scanning Electron Microscopy (FESEM, FEI Qunata 3D), which is also equipped to provide the cathodoluminescence (CL) spectra of the samples. Electrical properties were determined by Hall measurements at room temperature using Ecopia Hall measurement system (HMS 3000) and the surface chemistry was studied by X-ray photoelectron spectroscopy (XPS, Omicron). The as-deposited samples were directly inserted in the XPS chamber, without any surface treatment, to determine their native properties. Both non-monochromated Al K-$\alpha$ (1486.7) and Mg K-$\alpha$ (1253.6) were used as  X-ray sources.  After the initial studies, 0.8 KeV energy $Ar^{+}$ ion sputtering was employed for various times to remove the surface contaminants, followed by XPS core level and valence band analysis.

\section{Results and discussion}
Fig. 1 shows the plan view FESEM images of samples S1 to S4, which display the reducing surface roughness and porosity.  S1 is a highly porous random network of ‘nanowalls’, which will henceforth be referred to as nanowall network (NWN).  The width of the base of the tapered walls are a few hundred nanometres whereas the tips (apex) are less than 10 nm. S2 is a relatively flat film with high density of small pores, while sample S3 consists of large coalesced flat topped islands. S4 is the highly flat GaN epilayer. The surface roughness of the samples was quantified by atomic force microscopy (AFM).   

\begin{figure} [htbp]
\centering
\includegraphics[width=\textwidth]{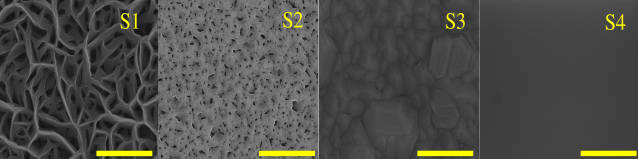}
\caption{The FESEM plan view images of S1-S4. The samples are progressively flatter. The scale  bar 
is 1 $\mu$m.}
\label{fig1}
\end{figure}

From the SEM image it is clear that since the pores of S1 are deep, the tapping mode AFM study may not provide an accurate value of roughness, but it is sufficient for comparison purposes. The RMS roughness of the samples from S1 to S4 were found to be 57.4, 19.9, 7.8 and 0.4 nm, respectively.  
The samples were cut into smaller pieces         ($\approx$ 1cm x 1cm), and indium metal contacts were made on one set of samples to study the electrical properties. Four probe Hall measurement, in van der Pauw geometry was carried out. The bulk electron concentration and resistivity of the samples are given in Table 1. All samples are unintentionally n-doped,with S1 and S2 having  bulk electron concentration of     $\backsim10^{20} cm^{-3}$ , while S3 has         $\backsim 10^{19} cm^{-3}$ and S4 has $\backsim 10^{17} cm^{-3}$ . Also it is seen that resistivities of S1 and S2 are about $10^{-3}$     $\Omega$-cm whereas S3 and S4 are about $10^{-1}$  $\Omega$-cm. 

\begin{table} [htbp]
\centering 
\caption{Electrical parameters from Hall measurement}
\begin{tabular}{ccc}
\hline
Sample name & Surface RMS roughness
(nm) & Bulk electron concentration \\
 & & (x $10^{18}$ $cm^{-3}$) \\
\hline
S1 & 57.4 & 244 \\
S2 & 19.9 & 122 \\
S3 & 7.82 & 9.7 \\
S4 & 0.38 & 0.2 \\
\hline
\end{tabular}
\label{table1}
\end{table}

A number of factors can be responsible for this unintentional doping, such as nitrogen vacancy ($V_{N}$), oxygen impurity ($O_{N}$)  \citep{reshchikov_behavior_1999} etc. A tiny amount of oxygen is observed, which may come from impurities in the source or by diffusion from the $Al_{2}O_{3}$ substrate.The natural propensity of GaN to be unintentionally n-doped has often been attributed to the formation of $V_{N}$ \citep{look_defect_1997}.
The XPS studies of the samples, were done by acquiring survey and core level spectra, without any surface treatment, to study the native surface. It was found that the GaN samples have oxygen and carbon contaminants on the surface due to handling, storage, and atmospheric exposure. The adventitious carbon is usually unreactive and does not react with gallium. The oxygen species can form stoichiometric ($Ga_{2}O_{3}$) or non-stoichiometric ($GaO_{x}$) gallium oxides or even gallium oxynitrides. For the analysis of XPS data, the adventitious carbon was assumed to be of hydrocarbon nature \citep{barr_nature_1995} and its binding energy (B.E.) was taken as 284.8 eV . For the analysis of core level spectra a Shirley background correction was employed and then deconvolution was performed using fitting parameters from the literature. The peak shape was taken to be Voigt, with Gaussian function of 80\% and Lorentzian of 20\%. Fig. 2. shows the deconvoluted Ga 3d spectra of samples S1-S4. The peak at 18.1 is due to the  Ga-Ga metallic bond \citep{shiozaki_formation_2007}. While the peak at ($19.3 \pm 0.2$) eV and the one at ($20.4 \pm 0.1$) eV have been assigned to Ga-N bond and Ga-O bond, respectively. \citep{fujishima_formation_2013}\citep{idczak_growth_2014} Prabhakaran et. al. \citep{prabhakaran_nature_1996} found the native oxide on GaN to be predominantly monoclinic $\beta-Ga_{2}O_{3}$  although they could  not rule out formation of oxynitrides. 

\begin{figure} [htbp]
\centering
\includegraphics[width=4cm]{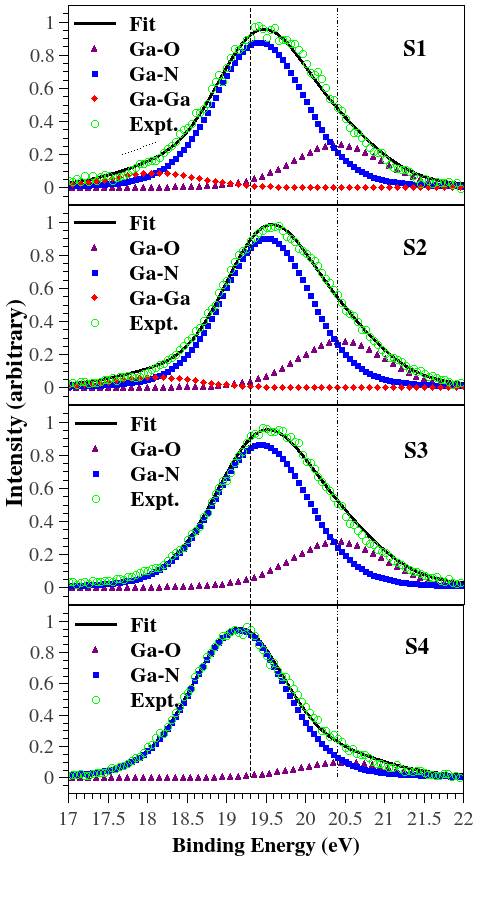}
\caption{The deconvoluted Ga 3d core level spectra of samples S1 - S4}
\label{fig2}
\end{figure}

The spectra in all cases predominantly show the Ga-N component, with Ga-O component occurring at higher B.E. value and the Ga-Ga component occurring at low B.E. value.  It can be observed in Fig. 2 that samples S1 and S2 contain Ga-Ga metallic bond which is absent in samples S3 and S4.To get the absolute surface composition of a sample from XPS, a completely clean surface is required. Various methods are employed to achieve completely clean surfaces, such as, ion sputtering \citep{kovac_surface_2002}, HCl cleaning \citep{king_cleaning_1998}, atomic hydrogen cleaning \citep{piper_clean_2005}, \emph{in situ} annealing, \citep{hattori_surface_2010} etc. Each of these methods have their own peculiar effects on the surface.The surface chemistry changes significantly after  prolonged $Ar^{+}$ sputtering, whereas without sputtering, the adsorbates mask the true composition.
 The thick GaN epilayer (S4) has been assumed to be stoichiometric GaN and relative compositions for sample S1-S3 have been estimated. A light 5 minutes sputtering (0.8 KeV, 50 $\mu$A) was employed on all the samples to remove some of the adsorbed contaminants. After that the peak intensities (area under peak after background subtraction) were measured and the elemental concentrations for Ga, N, O and C were estimated with appropriate atomic sensitivity factors (ASF), using the following relation:
\begin{equation}
\text{Concentration of } X = \frac{I_{X}/S_{X}}{\sum I_{i}/S_{i}}
\end{equation} 
where, $I_{i}$ and $S_{i}$ are intensity and ASF of i-th element.  	
In Fig. 3 the nitrogen to gallium concentration ratio has been plotted along with the bulk electron concentrations obtained from Hall measurements.  It can be seen from this figure that porous samples S1 and S2, have a Ga rich surface, with higher electron concentrations and lower resistivities (see Table 1). Sample S3 has the most Ga deficient surface and least electron concentration among the MBE grown samples studied.  

\begin{figure} [htbp]
\centering
\includegraphics[width=6cm]{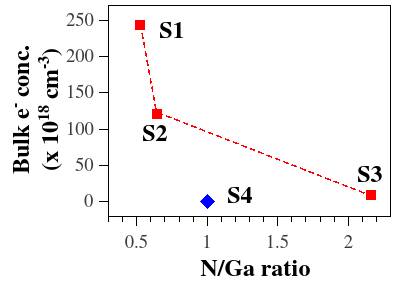}
\caption{Graph showing surface stoichiometry of 5 minute sputtered samples and their respective bulk electron concentration measured from Hall experiment.}
\label{fig3}
\end{figure}
 
A surface rich in gallium might mean either the surface is Ga terminated (N polar material), or that the N vacancies on a nonpolar/semipolar surface yield a termination consisting mostly of Ga atoms, whose dangling bonds reconstruct to minimize the high surface energy. Similar to that reported for InN, the thermodynamically stable GaN surface configuration is cation terminated with another cation adlayer on top of it, which can yield the Ga rich surface, \citep{segev_electronic_2007} forming an electron accumulation layer. Considering S3, a Ga deficient surface can indicate the presence of Ga vacancies, yielding the well known GaN yellow luminescence (YL) at around 2.2 eV ($\backsim$560 nm). \citep{neugebauer_gallium_1996},\citep{saarinen_observation_1997}

\begin{figure} [!h]
\centering
\includegraphics[width=4cm]{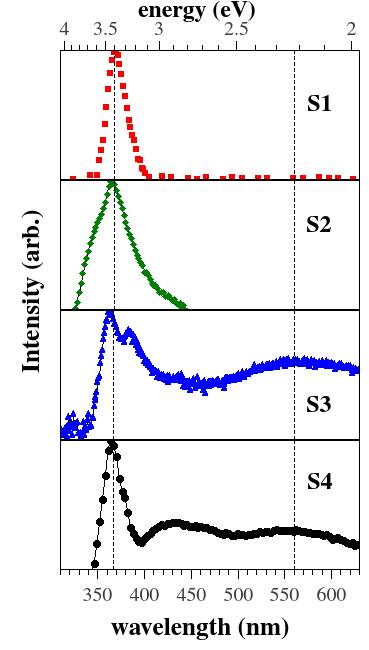}
\caption{Cathodoluminescence spectra of samples S1 to S4 obtained at room temperature with 20KV acceleration voltage and 4nA beam current.   
}
\label{fig4}
\end{figure}
 
To see the presence of such features, cathodoluminescence (CL) of samples S1-S4 was carried out at room temperature,with different acceleration voltages and electron beam  currents. Fig. 4 shows the CL spectra of  the samples with an acceleration voltage of 20 KV and a beam current of 4nA. All the samples yield a peak at  ($3.40 \pm 0.03$) eV, which is related to the band edge emission of GaN. It is clear that for the NBE emission, GaN epilayer (sample S4) has the lowest FWHM of 84 meV; followed by GaN NWN (sample S1) 104.4 meV, while S3 and S2 have higher widths of 111.3 meV and 134.4 meV, respectively. An important point to observe is
hat S1 has no defect related luminescence in the 400-600 nm region. Sample S2 has an unusually broad NBE peak with significant asymmetry. S3 has another peak around 3.2-3.25 eV and a broad peak around 2.2 eV which is the YL peak. Sample S4 has atleast two defect related peaks, one around 2.8 eV and one around 2.15 eV. The 2.8 eV blue luminescence (BL), $\backsim$ 3.25 eV UV luminescence (UVL) are  identified as defect related emissions in the literature, but the attribution to specific defects has been ambiguous. Some researchers \citep{shahedipour_investigation_2000},\citep{lin_surface_2002} ascribe the 2.8 BL to nitrogen vacancy or its complexes, eg. ($V_{N}$-H), while H.C.  Yang et. al. \citep{yang_nature_2000} relate it to the transition between $O_{N}$ donor level and ($V_{Ga} \text{-}  O_{N}$) complex deep level. Appearance of the 3.25 eV peak in the nominally undoped GaN sample is somewhat unusual, as it is  mostly seen in Mg:GaN materials, \citep{reshchikov_luminescence_2005}. Nevertheless this emission is sometimes explained to be due to nitrogen vacancies and incorporated hydrogen.  \citep{shahedipour_investigation_2000} 

To understand the variation of band structure due to the morphology of the films, we use XPS to locate the position of Fermi level, band bending and band filling. While the conduction band edge and surface Fermi level positions can be found from absorption (or luminescence) spectra and XPS valence band measurements, respectively, the position of Fermi level in the bulk of the sample is not easy to experimentally measure, unless the sample is degenerately doped. To estimate ($E_{F}$-$E_{V}$) in the bulk, a simple calculation was performed. \citep{streetman_solid_2006} For n-doped semiconductors, the electron concentration before doping ($n_{i}$) and after doping ($n_{0}$) are related as:\\
\begin{equation}
 n_{0} = n_{i} \text{  } exp(\dfrac{E_{F}-E_{i}}{K_{B}T}) 
\end{equation}

which can be rearranged to give

\begin{equation}
E_{F}-E_{i} = K_{B}T \text{  } ln(\dfrac{n_{0}}{n_{i}})
\end{equation}

Now,  $n_{i}$ is given as : \citep{noauthor_nsm_nodate}

\begin{equation}
n_{i} = \sqrt{N_{C}N_{V}} \text{  } exp(\dfrac{-E_{G}}{2K_{B}T})
\end{equation}

where, $N_{C}$ and $N_{V}$ are the conduction and valence band density of states, and other symbols have their usual meaning.  Further, these two quantities are given as : 

\begin{equation}
N_{C} = 2[\dfrac{2\pi m_{e} K_{B}T}{h^{2}}]^{\frac{3}{2}}
\end{equation}

\begin{equation}
N_{V} = 2[\dfrac{2\pi m_{h} K_{B}T}{h^{2}}]^{\frac{3}{2}}
\end{equation}

Using standard values and taking T=300 K, we get $N_{C}$ =  2.23x$10^{18}$ $cm^{-3}$  and  $N_{V}$ = 4.16x$10^{19}$ $cm^{-3}$ from which the intrinsic electron concentration is found to be:  $n_{i}$ = 3.87x$10^{-10}$ $cm^{-3}$. 

Then, by using the $n_{0}$ values from Table 1, we get the energy difference between bulk Fermi level and band midpoint ($E_{i}$)  for samples S1-S4 as:  1.78 eV, 1.76 eV, 1.70 eV and 1.60 eV, respectively, corresponding to decreasing n-doping. Thus, we observe that samples S1 and S2 are degenerately doped,which is also seen in UV-visible spectroscopy studies (not shown here) with absorption edge at $\backsim$ 3.5  eV, for sample S1.
To locate the position of Fermi level on the sample surface, a straight line was fit to the leading edge of the valence band spectra, whose intersection with the background gives the position of valence band maximum ($E_{V}$), with respect to $E_{F}$ = 0 eV binding energy. ($E_{F}$-$E_{V}$) values so found, are: 2.0 eV, 1.8 eV, 1.6 eV and 1.7 eV, respectively, for samples S1, S2, S3 and S4. The resulting band diagrams for S1 and S2 are shown in Fig 5. All the samples show an upward band bending of  1.44 eV, 1.63 eV, 1.8 eV and 1.6 eV, respectively. These values are also confirmed by measuring the separation between valence band maximum and Ga 3d core level, which is independent of band bending or any other effect. Using the ($E_{V}$ - $E_{Ga 3d}$ ) value of 17.76 eV, given by Waldrop and Grant, \citep{waldrop_measurement_1996} and the Ga 3d core level position, the position of surface Fermi level with respect to VBM,   ($E_{F}$-$E_{V}$) are found to be : 2 eV, 1.7 eV, 1.7 eV and 1.3 eV, for S1, S2, S3 and S4 respectively, which agree well with the first method.

\begin{figure} [!h]
\centering
\includegraphics[width=4cm]{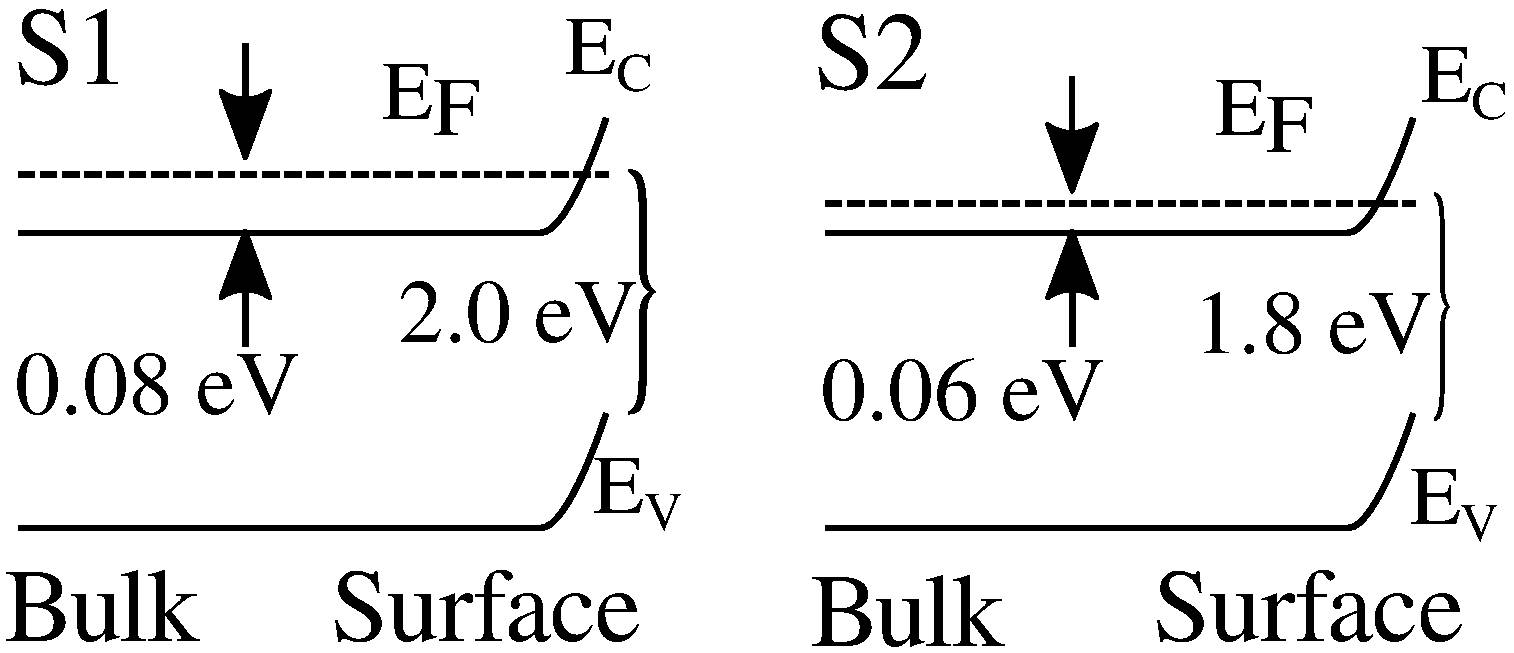}
\caption{The calculated band bending for samples S1 and S2. The band gap was found to be 3.4 eV from CL spectra. The surface Fermi level position was determined from XPS by linear fit of valence band leading edge. The bulk Fermi level position was estimated from band theory calculations.}
\label{fig5}
\end{figure}

Thus, the results show that the Fermi level gets pinned at ($1.8 \pm 0.2$) eV above VBM, for samples S1 and S2.  Earlier, the experimental work of Koçan et. al. \citep{kocan_surface_2002} showed that for GaN samples covered with a thin Ga adlayer, the Fermi level gets pinned 1.65 eV above VBM, and Van de Walle et. al. showed by DFT calculations \citep{segev_origins_2006} that for a metallic bilayer consisting of the terminating Ga atom and a Ga adatom, the Fermi level gets pinned at 1.8 eV above VBM for both polar and non-polar GaN surfaces.  Comparison of these literature values and our data suggests that a Ga adlayer exists at the surface of samples S1 and S2.  

Next, we study the effect of sputtering on the samples’ surface chemistry. $Ar^{+}$ sputtering is a very commonly used method to prepare clean surfaces. \citep{kovac_surface_2002} However, an inherent problem in that is the sputter yield, (average number of target atoms  sputtered per incident $Ar^{+}$ ion)  is not the same for all elements. Since the rate of nitrogen sputtering is more than that of gallium, the surface chemistry is significantly changed. Fig. 6 shows the core level Ga 3d spectra of S1 deconvoluted into Voigt components,  for the unsputtered sample and after 10 and 20 minutes of sputtering.
As it can be expected, with sputtering, the gallium oxide component reduces, while the metallic  Ga-Ga bond contribution increases. The calculated Ga/N absolute concentration ratio for unsputtered S1 sample and after 5, 10 and 20 minutes of sputtering, are: 0.3, 0.9,0.9 and 1.2, respectively, indicating that the surface becomes more metal rich with increasing $Ar^{+}$ sputtering. Such surface metallization with ion sputtering is also reported in the literature \citep{makowski_olefin_2011},\citep{carin_xps_1990}.

\begin{figure} [htbp]
\centering
\includegraphics[width=5cm]{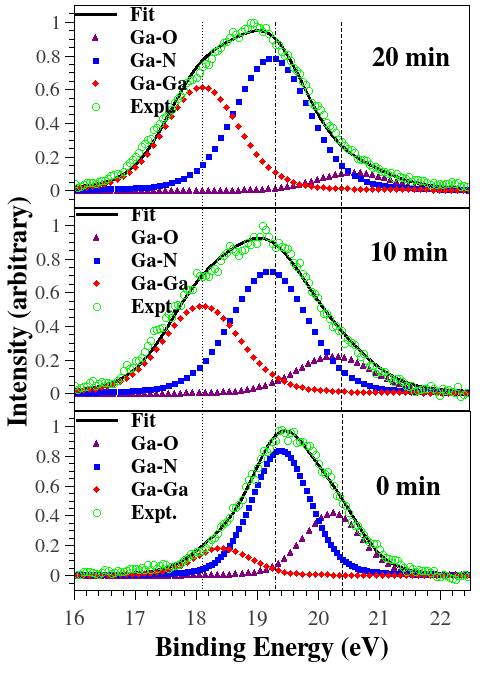}
\caption{Deconvoluted Ga 3d core level spectra of S1 at different sputtering times}
\label{fig6}
\end{figure}

Fig. 7 shows the normalized VB spectra of sample S1 at different sputtering times. Defects at semiconductor surfaces lead to a change in the DOS. \citep{cai_effect_2017} As VB spetcra reflects the total DOS of the semiconductor, the type and abundance of point defects can be estimated from VB spectra. A few observations can be made from Fig. 7.  The unsputtered VB spectra has two prominent features, one around 3-4 eV ($P_{A}$) and another around 8-9 eV ($P_{B}$). The low binding energy peak  originates from the hybridization of  Ga 4p and N 2p orbitals and is ‘p-like’. The higher binding energy peak is ‘s-like’ and is attributed to hybridization of Ga 4s and N 2p orbitals. The feature appearing in between these two peaks is attributed to adsorbates or mixed hybridized orbitals. \citep{mishra_surface_2015},\citep{mishra_pit_2015} It has been found from experimental studies that there is a direct correlation between relative intensities of $P_{A}$ and $P_{B}$ and the polarity of the sample.

\begin{figure} [htbp]
\centering
\includegraphics[width=6cm]{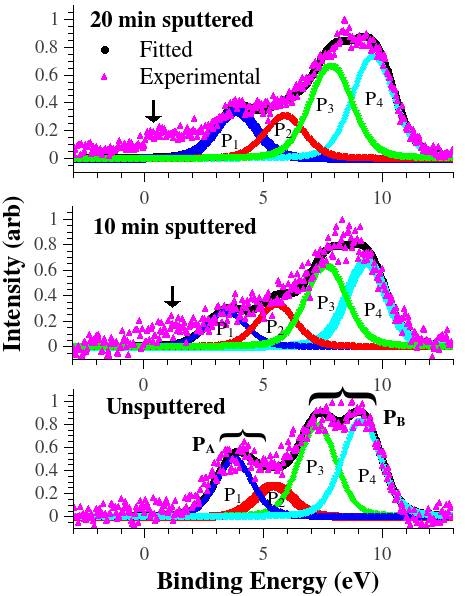}
\caption{Deconvoluted valence band spectra of S1 at different sputtering times. The arrow marks the build up of intensity near Fermi edge. }
\label{fig7}
\end{figure}

D. Skuridina et. al. \citep{skuridina_polarity_2013} showed that if the intensity of $P_{A}$ is higher than  $P_{B}$, the material is cation polar, whereas if the reverse occurs, it is N polar. So, from the Fig. 7, it appears that S1 is N polar.  After this initial visual analysis, the VB spectra was fitted by 4 Voigt peaks each, in the way described earlier. The peaks at 5.4 eV and 9.1 eV are assigned to Ga 4s – N 2p hybridized orbitals. The peak at 3.8 eV is assigned to Ga 4s – N 2p hybridized orbitals.\citep{magnuson_electronic_2010}

\begin{table} [htbp]
\centering
\caption{Assignment of deconvoluted peaks of valence band spectra}
\begin{tabular}{ccc}
\hline
Peak & Binding energy (eV) & Origin \\
\hline
P1 & 3.8 & Ga 4p - N 2p \\
P2 & 5.4 & Ga 4s – N 2p*\\
P3 & 7.3 & Mixed hybrid orbitals \\
P4 & 9.1 & Ga 4s – N 2p \\
\hline
\end{tabular}
\end{table}

The peak $P_{3}$, as mentioned before, is generally attributed to be either adsorbate related or  due to mixed hybrid orbitals. Since this peak intensity did not change with sputtering, we discount the role of adsorbates, and attribute it to mixed hybrid orbitals. A prominent observation that can be made from Fig. 7 is that a significant intensity can be seen gradually building up near the Fermi level position with sputtering, which is a characteristic of surface metallization as also seen in the Ga 3d core level changes in Fig. 6.

\section{Conclusion}
In conclusion, MBE grown GaN samples of different morphologies were studied. It was found that GaN nanowall network has high conductivity and no defect related luminescence. From XPS measurements and theoretical calculations it was found that the Fermi level gets pinned ($1.8 \pm 0.2$) eV for the nanowall, which indicates that there is a Ga metallic adlayer present on its surface. The flatter samples showed lower conductivity and defect related yellow luminescence due to Ga vacancies at their surfaces. It was found that $Ar^{+}$ sputtering, leads to surface metallization as confirmed from XPS core level and valence band measurements. This study helps gain an insight into the correlation between morphology and surface chemistry that is useful in employing such systems to fabricate devices.

\section{Acknowledgement}
The authors thank Prof. C. N. R. Rao for his support and guidance. AC acknowledges DST for a Senior Research Fellowship.

\bibliographystyle{unsrtnat}
\bibliography{Abhijit_Chatterjee_morphology_surface_GaN_arxiv}

\end{document}